\documentstyle[preprint,aps]{revtex}
\preprint{\vbox{  
\hbox{IFT-P.020/99}   
\hbox{February 1999} 
}}  
\begin{document}  
\draft 
\title{
Comment on ``Majoron emitting neutrinoless double beta decay in the 
electroweak chiral gauge extensions''}
\author{J. C. Montero, C. A. de S. Pires and V. Pleitez} 
\address{
Instituto de F\'\i sica  Te\'orica\\ 
Universidade  Estadual Paulista\\
Rua Pamplona, 145\\ 
01405-900-- S\~ao Paulo, SP\\ 
Brazil } 
\date{\today}
\maketitle 
\begin{abstract}  
We point out that if the majoron-like scheme is implemented within a 331 model,
there must exist at least three different mass scales for the scalar vacuum 
expectation values in the model.
\end{abstract}
\narrowtext   
In a recent paper by Pisano and Sharma~\cite{ps} the majoron scheme was 
implemented in a 331 model~\cite{331}.
In that paper two different scales of vacuum expectation values (VEV) in the 
scalar sector have been considered: one related with the electroweak symmetry 
breaking and the other one with the $SU(3)$ breaking. 
Here we show that the model is consistent with the experimental value of the 
$\rho$ parameter only if three mass scales are introduced.

It is a well known fact that Higgs triplets under the standard 
$SU(2)_L\otimes U(1)_Y$ gauge group have to have vacuum 
expectation values which are smaller than the electroweak scale in order 
to not spoil the agreement between the theoretical and the experimental value of 
the electroweak $\rho$ parameter ($\rho=M_Z/M_W c_W$)~\cite{r1,r2,r3}. 
This is due to the fact that triplets and doublets give different 
contributions to the $W^\pm$ and $Z$-boson masses. 
This result does not depend on the hypercharge of the Higgs triplet. 
For instance, a Higgs doublet and a Higgs triplet with $Y=2$
with spontaneous (Majoron Model~\cite{r1}) or explicit (non-Majoron 
Model~\cite{r2,r3}) lepton number violation give:
\begin{equation}
M_W^2=\frac{g^2}{4}(v_D^2+2v_T^2), \;
M_Z^2=\frac{g^2}{4c^2_W}\left(v_D^2+4v_T^2\right)
\label{wzmass}
\end{equation}
where $v_D$ and $v_T$ denote the VEVs of the doublet and the triplet, 
respectively. 
Notice that the condition $v_D=v_T$ violates the $\rho=1$ condition.
We can not even use $v_D^2+2v_T^2=(\mbox{246 GeV})^2$. The only way to avoid 
this problem is that $v_T \leq 5.5$ GeV (if $v_D=246$ GeV) using the present 
experimental value for the $\rho$-parameter (see below).
Thus, we see that these sort of models have two different mass scales:  
$v_T$ and  $v_D$.

Next, let us consider a similar situation in the context of
the 331 model~\cite{331}. In that model in order to give mass to all the 
fermions it is necessary to introduce three Higgs triplets and a Higgs sextet. 
Two of the triplets and the sextet have the 
neutral component in a doublet of the subgroup $SU(2)$; we denote the respective 
VEVs by $v_\eta, v_\rho$ and $v_{DS}$. The other triplet has its neutral 
component transforming as a singlet under $SU(2)$ and the sextet has another 
neutral field transforming as a triplet under $SU(2)$. Let us denote
their respective VEVs by $v_\chi$ (the VEV which is in control of the $SU(3)$ 
breaking) and $v_{TS}$. 
The $W^\pm$ and $Z$-boson masses, neglecting terms of the order 
$v_iv_j/v^2_\chi$ ($i,j=\eta,\rho,DS,DT$), 
are given by:
\begin{equation}
M_W^2\approx\frac{g^2}{4}(v_\eta^2 + v_\rho^2+2v_{DS}^2 + 4 v_{TS}^2),
\label{wmass}
\end{equation}
and
\begin{equation}
M_Z^2\approx \frac{g^2}{4c^2_W}\left( v^2_\eta+v^2_\rho+2v^2_{DS}+ 8 
v_{TS}^2\right),
\label{zmass}
\end{equation}
respectively.

We see from Eqs.(\ref{wmass}) and (\ref{zmass}) that as in the case of the 
$SU(2)_L\otimes U(1)_Y$ model given in Eq.~(\ref{wzmass}), the triplet 
$v_{TS}$ contributes in a different way to the $W^\pm$ and $Z$-boson masses. 
We can estimate the order of magnitude of $v_{TS}$ by assuming that
$v_\eta\approx v_\rho\approx v_{DS}\equiv \tilde{v}$ and
using the experimental value of the $\rho$-parameter: 
$\rho=0.9998\pm0.0008$~\cite{pdg}. 
From Eqs.~(\ref{wmass}) and (\ref{zmass}) we have
\begin{equation}
\rho= \frac{1+r}{1+2r},\quad \sqrt{r}=\frac{v_{TS}}{\tilde{v}},
\label{ulti}
\end{equation}
which implies the upper limit for $r\leq0.001$ or $v_{TS}\leq3.89$ GeV for
$\tilde{v}^2=(246/2)^2\; {\rm GeV}^2$.
If we make $v_{TS}=v_\eta=v_\rho=v_{DS}$ as it has been done in Ref.~\cite{ps} 
we violate the $\rho=1$ condition (it gives $\rho=2/3$).
In conclusion, the model must have at least three different mass scales in the 
scalar vacuum expectation values: $v_{TS},\, \tilde{v}$ and $v_\chi$.

\acknowledgments 
This work was supported by Funda\c{c}\~ao de Amparo \`a Pesquisa
do Estado de S\~ao Paulo (FAPESP), Conselho Nacional de 
Ci\^encia e Tecnologia (CNPq) and by Programa de Apoio a
N\'ucleos de Excel\^encia (PRONEX). One of us (CP) would like to thank
Coordenadoria de Aperfei\c coamento de Pessoal de N\'\i vel Superior (CAPES) 
for financial support.

\end{document}